# Hierarchical Triple-Modular Redundancy (H-TMR) Network For Digital Systems

B. Baykant ALAGOZ

*Abstract: Hierarchical application of Triple-Modular Redundancy (TMR) increases fault tolerance of digital Integrated Circuit (IC). In this paper, a simple probabilistic model was proposed for analysis of fault masking performance of hierarchical TMR networks. Performance improvements obtained by second order TMR network were theoretically compared with first order TMR network.*

## 1. Introduction:

Fault tolerance becomes substantial design criteria for the applications where the reliability of hardware was crucial. Medical, military and long-range missions are such applications that the fault tolerance of hardware became key issue. Faults affecting the system can be permanent nature resulting from decay in materials or transient nature resulting from extreme working condition such as heavy electromagnetic disturbance. Triple Modular Redundancy (TMR) was the most applied fault masking technique for fault tolerance of software or hardware systems.

TMR technique uses three implementation of the same function (redundant modules) and outputs of these modules are voted by a voting mechanism. The most basic voting algorithm is majority voting, where voter algorithm selects the most common output. Truth table of conventional bit-by-bit majority voter logic circuit was given in Table 1.

**Table 1**. Truth table of fault masking block

| $y_1$ | $y_2$ | $y_3$ | $y$ | Alarm |
|---|---|---|---|---|
| 0 | 0 | 0 | 0 | 0 |
| 1 | 0 | 0 | 0 | 1 |
| 0 | 1 | 0 | 0 | 1 |
| 1 | 1 | 0 | 1 | 1 |
| 0 | 0 | 1 | 0 | 1 |
| 1 | 0 | 1 | 1 | 1 |
| 0 | 1 | 1 | 1 | 1 |
| 1 | 1 | 1 | 1 | 0 |

Logic equation for majority voting circuit of TMR technique was written as following,

$$y = y_1.y_2.\bar{y}_3 + y_1.\bar{y}_2.y_3 + \bar{y}_1.y_2.y_3 + y_1.y_2.y_3 \quad (1)$$

Logic equation of fault indication signal that monitors weak consensus at redundant modules can be expressed as,

$$Alarm = (y_1 + y_2 + y_3).(\bar{y}_1 + \bar{y}_2 + \bar{y}_3) \qquad (2)$$

TMR masks effect of faults before spreading trough the rest of system. Triple modular redundancy (TMR) has been widely used in highly reliable applications. Lots of studies were done to increase fault-masking performance of TMR in literature. Word-Voter technique was developed and its advantages over the bit-by-bit voting schemes was shown.[1] Dynamic fault-tolerant system was suggested and reliability of the model was shown.[2] In order to increase reliability, self-diagnosis elements and voting algorithms were cooperated.[3] Experimental results show that the appropriate use of diagnosis in a fault masking system enables the voter to select more correct results than results of majority voting alone. Adaptive voter algorithm using confidence value was shown to improve reliability of the system.[4] TMR techniques have been researched to gain sufficient fault tolerance to digital systems against single error, multiple-bit errors, transient or permanent errors. [5]-[8] In this paper, we focused on improving reliability by hierarchically application of conventional bit by bit TMR to high volume digital systems such as digital ICs. For this purposes, we suggested TMR flip-flop concept to simplify implementation of hierarchical TMR network on the today's digital ICs. Data paths in digital IC are considered as redundant modules of TMR and three replicas of every data paths connected to a TMR Flip flop, which is composed of a majority voting circuited defined by Equation 1, a fault indication logic defined by Equation 2 and a regular flip flop as a register. Figure 1 demonstrates TMR application to digital ICs.

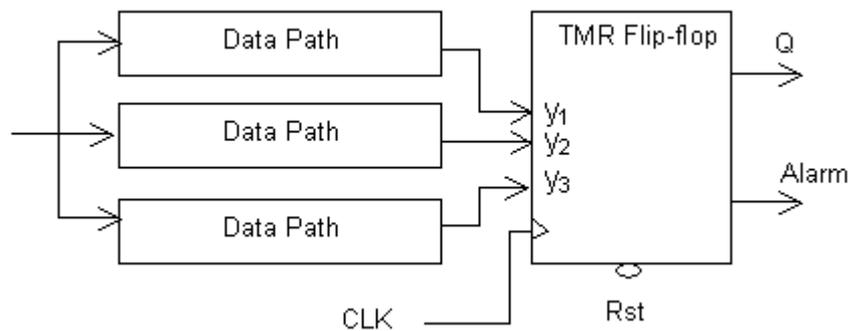

**Figure 1.** TMR application concept for Digital IC Design

TMR flip-flops concept facilitates TMR deployment in high volume logic design process. Figure 2 illustrates TMR flip flop inner structure.

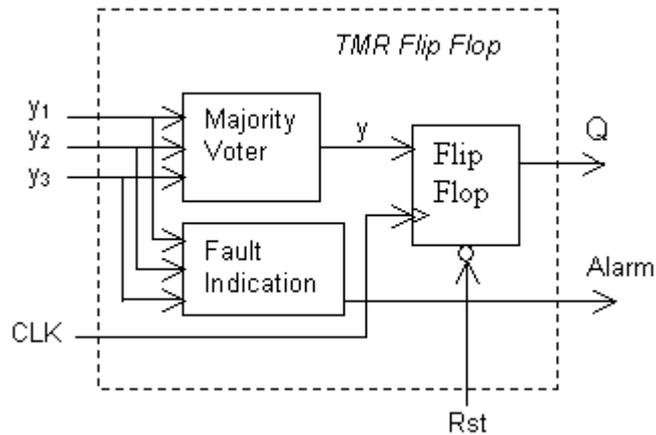

**Figure 2.** TMR application concept for Digital IC Design

TMR flip-flops constitute a TMR network on digital logic system. When redundant modules of TMR contain other TMR flip-flops, a TMR support hierarchy is emerged in TMR network as result of one TMR flip-flops support covering the other TMR flip-flops. In the Figure 6, we have seen second order TMR supports. In the Figure, last TMR flips-flop supports first three TMR flip-flops and it grants second order in TMR support hierarchy. We have theoretically showed that increasing hierarchy in TMR network would significantly improve fault tolerance of the digital hardware.

## 2. Performance Analysis Of Hierarchical TMR Network By Error Probability Models:

Error probability model was developed for the analysis of fault masking performance of TMR network in a faulty digital system. First, error probability and accordingly error reduction rate of first order TMR support was investigated. Then error probability and error reduction rate analysis was made for second order TMR support. Finally, analysis was enlarged for higher order TMR network [9].

**2.1. Error Probability Model for First Order TMR Support:**

Lets denote error probability of the logic function $P_f$, and probability of being fault-free $P_n$, one can write down following equation,

$$P_n = 1 - P_f \qquad (3)$$

Faulty status for one-bit TMR and corresponding probabilities were listed in Table 2. In this table, N represents fault-free state and F represents faulty state of logic functions.

**Table 2**. Error probability of one-bit TMR

| $y_1$ | $y_2$ | $y_3$ | Selected Modules | Error Status at Output y | Error Probability |
|---|---|---|---|---|---|
| N | N | N | $y_1, y_2, y_3$ | None | 0 |
| F | N | N | $y_2, y_3$ | None | 0 |
| N | F | N | $y_1, y_3$ | None | 0 |
| N | N | F | $y_1, y_2$ | None | 0 |
| N | F | F | $y_2, y_3$ | Error | $P_n.P_f.P_f$ |
| F | N | F | $y_1, y_3$ | Error | $P_f.P_n.P_f$ |
| F | F | N | $y_1, y_2$ | Error | $P_f.P_f.P_n$ |
| F | F | F | $y_1, y_2, y_3$ | Error | $P_n.P_f.P_f$ |

We assumed that every fault in functionality would bring out an error at the output of modules. When the TMR flip-flops were assumed to fault-free, error probability of the system supported by single TMR would be sum of fault probabilities seen in Table 2.

$$Pe(P_f) = 3P_f^2 - 2P_f^3 \qquad (4)$$

*Pe* denotes error probability of system supported by TMR and $P_f$ is error probability of logic functions (redundant modules). When the TMR flip-flops were assumed to fault-free, error probability of the TMR can be written as,

$$Pem(P_f, P_{fmb}) = P_f.P_{fmb} + Pe(P_f).(1 - P_{fmb}) \qquad (5)$$

*Pem* denotes error probability of TMR when TMR flip-flops have $P_{fmb}$ error probability. When error probability of the TMR flip-flops comes closer to zero, error probability of system approximates to Equation 4.

In the Figure 3, characteristics of the error probability functions *Pe*, $P_{em}$ and $P_f$ were drawn for first order TMR network that does not contains any TMR flip-flops. In drawing of the $P_{em}$, it was assumed that $P_{fmb} = P_f$ and all redundant modules have $P_f$ fault probability. In the error probability characteristics, a weakness of TMR is taken attention. When error probability of redundant modules ($P_f$) exceeds the 0.5 probability, common errors (common mode failures) become majority at function's outputs and they fail TMR to perform fault masking, appropriately. Unfortunately, error probability of TMR implemented system becomes higher then the error probability of the redundant modules itself, when the error probability of modules exceeds 0.5 probability. In case of two modules being fault-free and one module being faulty, which was said to single error mode, TMR was shown to mask all errors.(see Figure 11.) In single error case, TMR provides perfect fault tolerance.

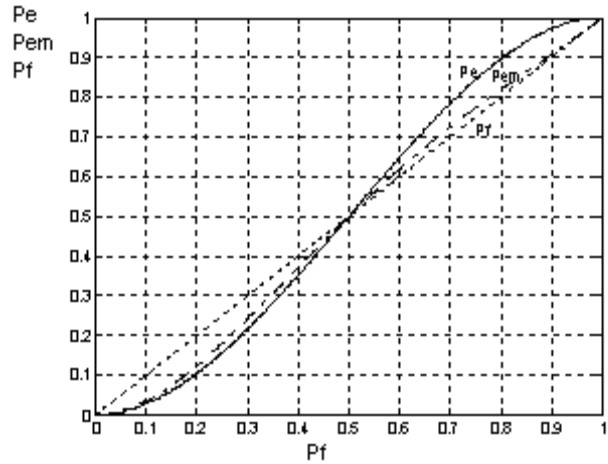

**Figure 3**. Error probability functions of TMR

In order to better express fault-masking performance of the TMR, lets define fault reduction rates as follows,

$$\text{Re} = \log \frac{P_f}{Pe} \qquad (6)$$

$$\text{Re}m = \log \frac{P_f}{Pem} \qquad (7)$$

In Figure 4, under assumption of $P_{fmb} = P_f$, error reduction rate of first order TRM system was drawn respect to $P_f$.

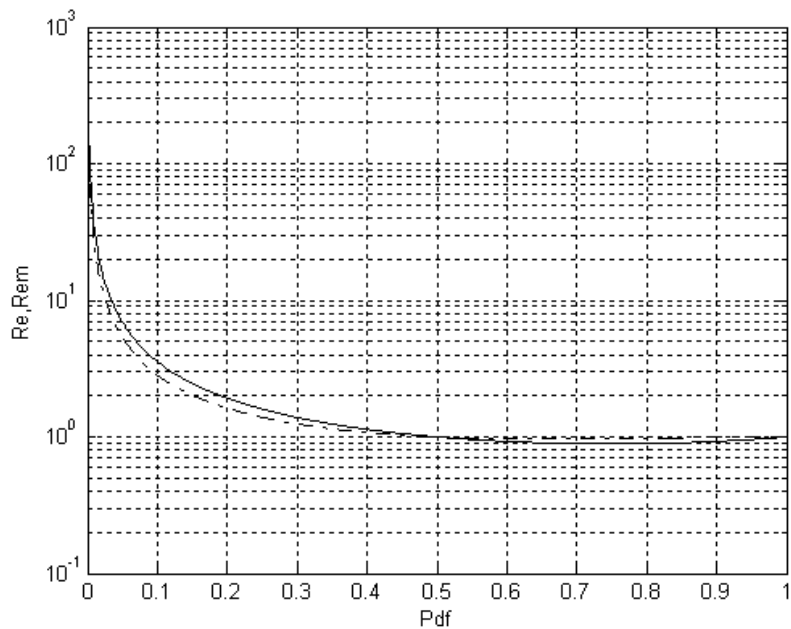

**Figure 4**. Error Reduction Rate of TMR. (*Re*-dashed line, *Rem*-solid line)

As seen Figure 4, TRM system provides remarkable reduction in error probability when logic function error probability $P_f$ is lower than probability of 0.1.

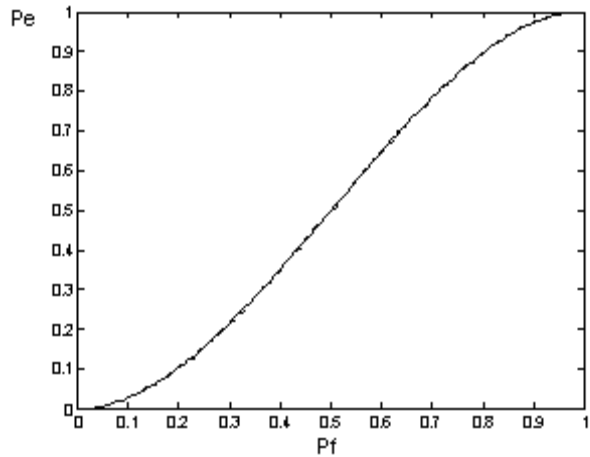

**Figure 5.a**. Error probability results

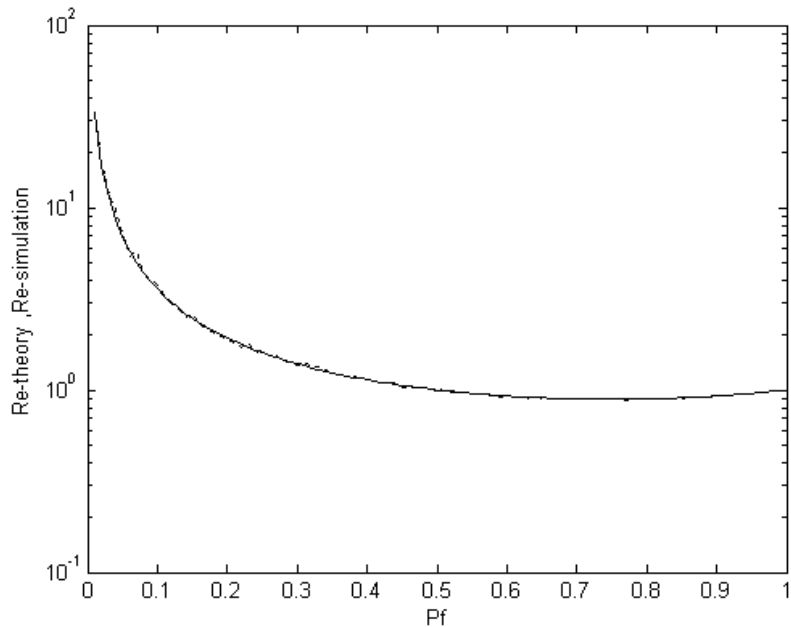

**Figure 5.b**. Error Reduction Rate results
**Figure 5**. Simulation results (dashed line) and Theoretical results (solid line)

In Figure 5, theoretical results obtained by Equation 4 were compared by simulation results of the TMR in digital system. In the simulation of TMR system, uniform distributed random errors were inserted into outputs of the three redundant logic functions. Totally 10.000 bits were processed by TMR for each error probability value ranging from 0.0 to 1.0. It is clearly seen from Figure 5, that theoretical error probability model was consistent with simulation results.

## 2.2. Error probability Model for Second Order TMR Support and Further Orders TMR Networks:

When a data path contains TMR flip-flops, the TMR flip-flops, supporting this data path, will constitutes second order TMR support in hierarchy of TMR network. Figure 6 illustrates second order TMR support implementation by TMR flip-flops.

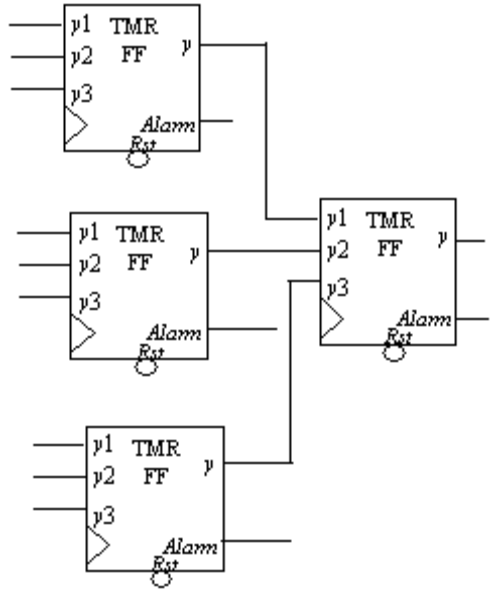

**Figure 6**. Representation of the second order TMR in hierarchy

When data paths includes second order TMR implementations, the TMR flip-flops supporting this data paths will be considered as third order TMR and so further. Error probability analysis of cascade connected higher order TMR supports as seen in Figure 6 was done by using error probability equation produced for lower order TMR supports.

Lets denote error probabilities of the $j$ order TRM $Pe_j$. Fault probability equations for $j$ order of TMR implementation can be written depending on error probability function of $j$-1 order, because, $j$-1 order TMR output directly feeds the $j$ order TMR system. For the error probability of j order TMR deployment, we can write following equations,

$$Pe_j(Pe_{j-1}) = 3Pe_{j-1}^2 - 2Pe_{j-1}^3 \qquad (8)$$

$$Pem_j(Pem_{j-1}, P_{fmb}) = Pem_{j-1}.P_{fmb} + Pe_j(Pe_{j-1}).(1 - P_{fmb}) \qquad (9)$$

Lets consider error probability of the second order of TMR seen in Figure 6 and write down error probability equations by setting $j$ to 2,

$$Pe_2(Pe_1) = 3Pe_1^2 - 2Pe_1^3 \qquad (10)$$

$$Pem_2(Pem_1, P_{fmb}) = Pem_1.P_{fmb} + Pe_2(Pe_1).(1 - P_{fmb}) \qquad (11)$$

Equation 4 is used for calculation of $Pe_1$ and equation 5 is used for calculation of $Pem_1$.

Lets draw error probability characteristics for the first and second order TRM respect to redundant module error probability $P_f$ in Figure 7, under assumption of $P_{fmb} = 0$.

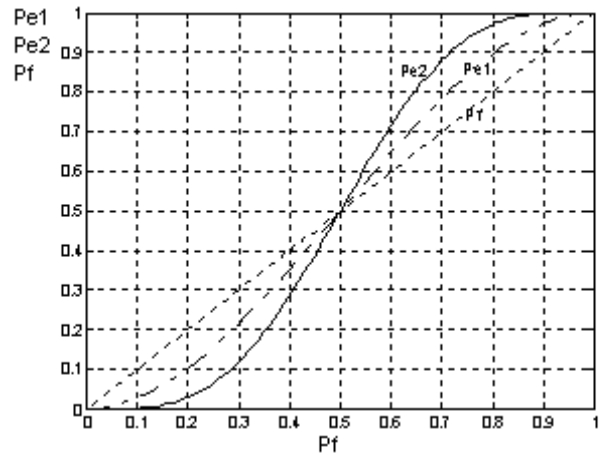

**Figure 7**. Error probability characteristics of the first (dashed line) and the second (solid line) order TMR

And fault reduction rate is correspondingly written as following,

$$\mathrm{Re}_2 = \log \frac{P_f}{Pe_2} \qquad (12)$$

$$\mathrm{Re}\,m_2 = \log \frac{P_f}{Pem_2} \qquad (13)$$

When the error reduction characteristics seen in Figure 8 is considered, it is seen that the second order TMR has much better fault reduction rate than the first order TMR in low error probability region ($P_f < 0.1$), as a result of increasing number of redundant modules involving voting.

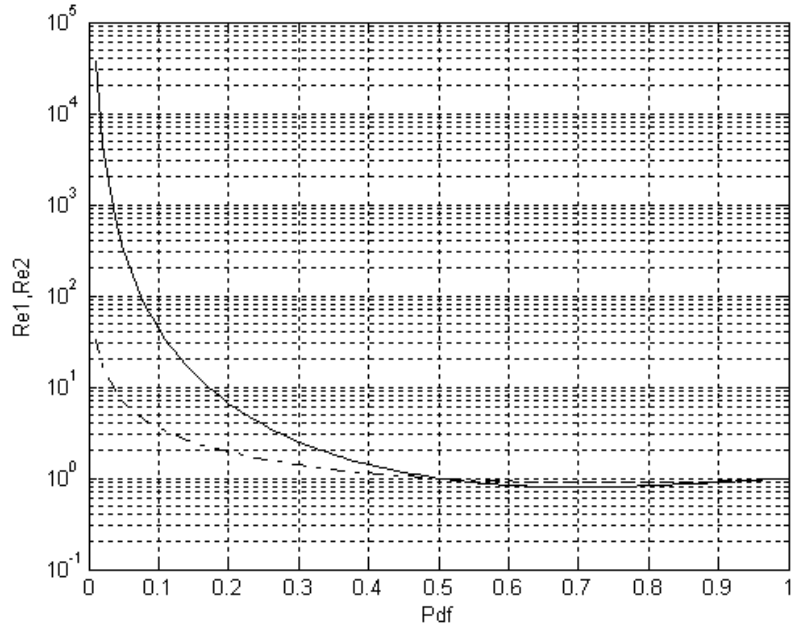

**Figure 8**. Error Reduction Rate of the second order TMR (Re$_2$- solid line) and the first order TMR (Re$_1$- dashed line)

While the error probability of logic function was lower than 0.1, approximately more than ten times higher reduction in error probability in the second order TMR was obtained compared to first order TMR and more than forty time higher reduction in probability was obtained compared to error probability of logic function without TMR support. For better comparisons between fault-masking performances of TMR hierarchy, Table 3 was constructed according to error probably.

**Table 3**. Number of operation resulting one error

| Error Probability | Number Of Operation Resulting One Error At Output Of System | | |
|---|---|---|---|
| | Data Path (Module) | First order TMR | Second Order TMR |
| 0.001 | $10^3$ | $3.3 \, 10^5$ | $3.7 \, 10^{10}$ |
| 0.01 | 100 | $3.3 \, 10^3$ | $3.7 \, 10^6$ |
| 0.1 | 10 | 35 | 433 |
| 0.3 | 3 | 4 | 8 |
| 0.5 | 2 | 2 | 2 |

**2.3. Self-diagnosis By Fault Alarm Indicator Signal:**
Fault alarm indicator signal of TMR Flip-flop can be used for self-diagnosis of functional failure that leads errors at the output. Logic equation of fault alarm indicator signal was given at (2). Logic one of the signal indicates incoherence at the output of modules and infers a faulty case in hardware. When all three modules are faulty, fault alarm indicator signal can't detect it, since there isn't any incoherence at outputs. (see Table 2)

Alarm indicator signals of TMR flip-flops can be monitored to estimate system health. *Alarm* signals from TMR flip-flops can be count by a Fault Status Counter to periodically checking quality of hardware. For the re-configurable hardware such as FPGA, *Alarm* signal can be used for triggering repair process.

## 3. Simulation Strategy And Results

For the TMR simulation, three different faulty module scenarios were simulated. These test scenarios are written down,

$$TestSenario_{TMR} = \{(N,N,F),(N,F,F),(F,F,F)\} \quad (14)$$

Here, *F* represents faulty modules and *N* represents fault-free modules. Elements of test scenario represent a test session with corresponding modules type that is either faulty (F) or fault-free (N). First scenario *(N,N,F)* is for one faulty and two fault-free modules. In this test session, performance of TMR network against single mode error was observed. In Figure 9, error reduction rates in *(N,N,F)* scenario was demonstrated for the first order and second order TMR. It masked all single mode errors.

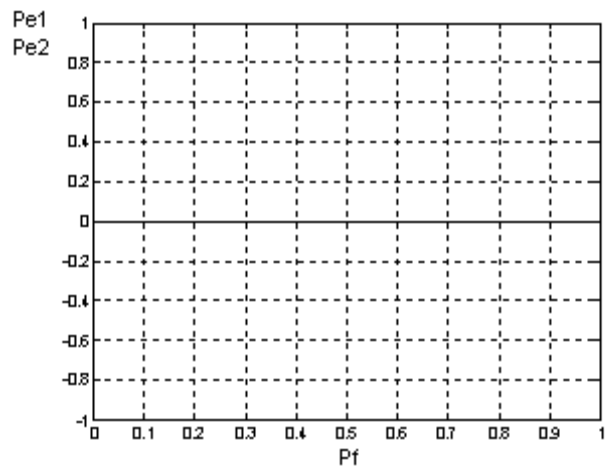

**Figure 9**. Error probability characteristics of the first (dashed line) and the second (solid line) order TMR in test scenario of *(N,N,F)*

Second scenario *(N,F,F)* is for two faulty and one fault-free modules. In this test session, performance of TMR support against both single mode error and common mode error were observed. In Figure 10, error reduction rates in *(N,N,F)* scenario was demonstrated for the first order and second order TMR.

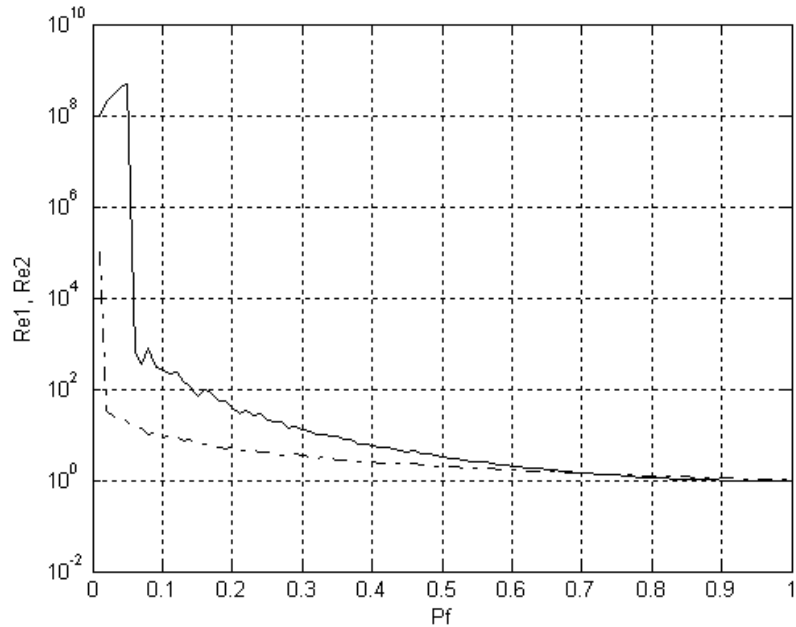

**Figure 10**. Error Reduction Rate of the second order TMR ($Re_2$- solid line) and the first order TMR ($Re_1$- dashed line) in test scenario of *(N,F,F)*

Third scenario *(F,F,F)* is for three faulty modules. In this test session, performance of TMR support against single mode error, common mode error and fully error were observed. Figure 5 had been drawn for the *(F,F,F)* scenario in previous section. In Figure 11, error reduction rates in *(F,F,F)* scenario was demonstrated for the first order and second order TMR.

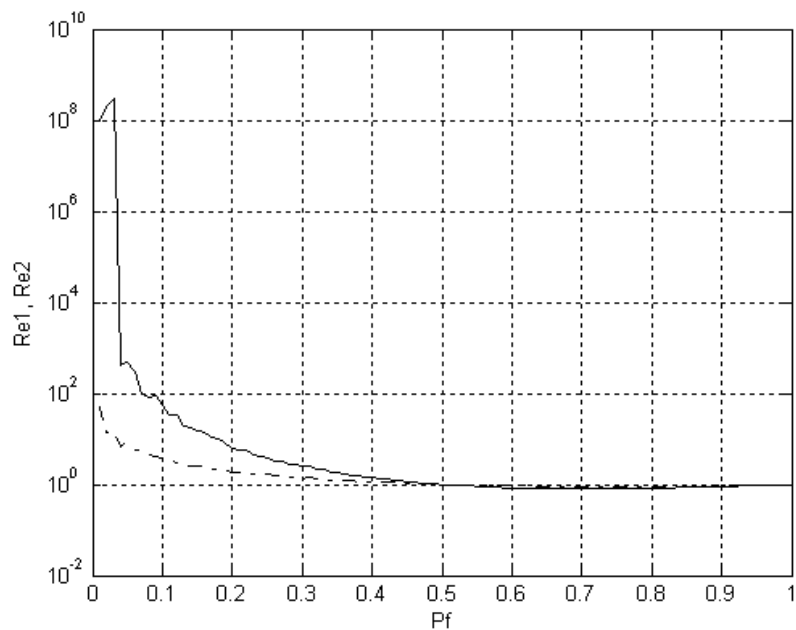

**Figure 11**. Error Reduction Rate of the second order TMR ($Re_2$- solid line) and the first order TMR ($Re_1$- dashed line) in test scenario of *(F,F,F)*

Faulty module in simulations was composed of a mutiplexer, a fault-free module and uniform distributed random error generator, which responsible to production error fitting error probability requirement ranging from 0.0 to 1.0.

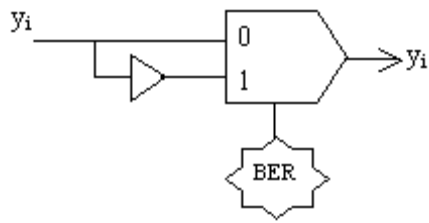

**Figure 12**. Faulty module model

BER block in Figure 12 randomly switches from $y_i$ to inversion of $y_i$ whenever error insertion is needed to satisfy error probability.

*Some observations from theoretical calculation and simulation results;*

1-TMR network exhibits very high error reduction rates at very low error probabilities. Especially, it provides very effective fault masking against transient random error at low error probability region, even if all module was faulty. Besides, it was known that TMR masks all errors if two out of three modules were fault-free. (see Figure 9)  When two modules are faulty, it considerably reduces error probability at low error probability of the redundant modules. (see Figure 10)

2-Second order TMR has much better error reduction performance at very low error probability regimes. Figure 13 shows error reduction rates at very low error probability region where error probability of modules is lower then $10^{-4}$.

3-Second order TMR requires totally 9 redundant logic functions. First order TMR requires 3 redundant logic functions. Figure 14 shows error reduction performance per redundant module at very low error probability region. Second order TMR showed superior error reduction performance per redundant modules compared to first order TMR. Increasing order of TMR network results better fault masking performance per modules.

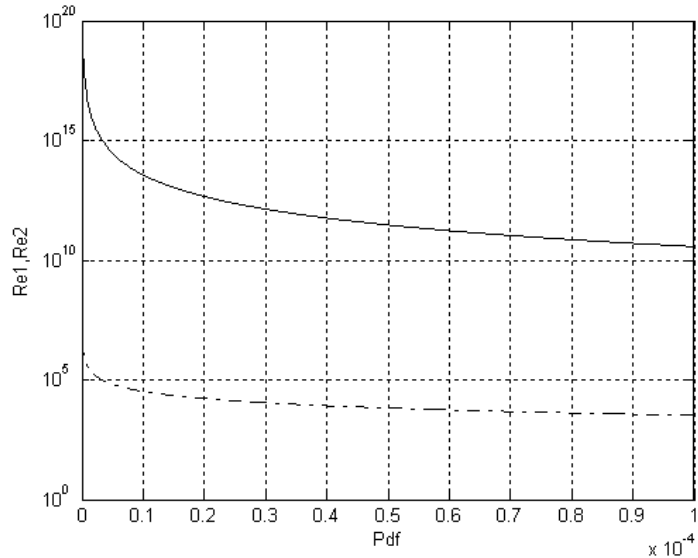

**Figure 13**. Error Reduction Rate of the second order TMR ($Re_2$- solid line) and the first order TMR ($Re_1$- dashed line) at very low error probability region ($Pdf < 10^{-4}$)

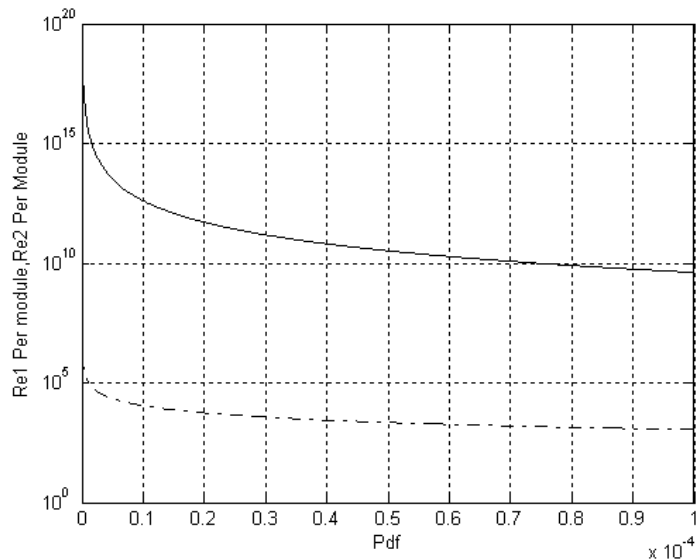

**Figure 14**. Error Reduction Rate per redundant module for the second order TMR ($Re_2$- solid line) and the first order TMR ($Re_1$- dashed line) at very low error probability region ($Pdf < 10^{-4}$)

## 4. Conclusions:

In this study, a design and analysis method for hierarchical TMR network distributed into high volume logic was theoretically researched. This research showed us that at the very low error probability of logic functions, high order TMR network implementation in digital system would remarkably reduce error probability of digital systems. It will gain significant fault tolerance and help perform its function correctly under transient or permanent failures.

For the further study, experimental studies of hierarchical TMR support for today's digital IC technologies (ASIC, FPGA) should be done and its fault tolerance performance under hazardous and extreme working conditions should be validated by experiments. We expect to see that TMR support would increase digital IC lifetime and also it would be slightly broadened ranges of working conditions for the digital ICs.

## Proposition:

Error probability of digital system supported by higher order TMR network is smaller than error probability of digital system supported by lower order TMR network. ($Pe_j < Pe_{j-1}$)

## Proof:

Lets write $Pe_{j-1}$ depending onto $Pe_{j-2}$ as following,

$$Pe_{j-1}(Pe_{j-2}) = 3 \cdot Pe_{j-2}^2 - 2 \cdot Pe_{j-2}^3 \qquad (15)$$

And write $Pe_j$ depending onto $Pe_{j-1}$ as following,

$$Pe_j(Pe_{j-1}) = 3 \cdot Pe_{j-1}^2 - 2 \cdot Pe_{j-1}^3 \qquad (16)$$

Lets arrange equation 16 by using equation 15,

$$Pe_j = 27 \cdot Pe_{j-2}^4 - 18 \cdot Pe_{j-2}^5 - 42 \cdot Pe_{j-2}^6 - 72 \cdot Pe_{j-2}^7 + 48 \cdot Pe_{j-2}^8 - 16 \cdot Pe_{j-2}^9 \qquad (17)$$

In order to better comparison, one can express $\dfrac{Pe_{j-1}}{Pe_{j-2}^2}$ and $\dfrac{Pe_j}{Pe_{j-2}^2}$.

$$\frac{Pe_{j-1}}{Pe_{j-2}^2} = 3 - 2 \cdot Pe_{j-2} \qquad (18)$$

$$\frac{Pe_j}{Pe_{j-2}^2} = 27 \cdot Pe_{j-2}^2 - 18 \cdot Pe_{j-2}^3 - 42 \cdot Pe_{j-2}^4 - 72 \cdot Pe_{j-2}^5 + 48 \cdot Pe_{j-2}^6 - 16 \cdot Pe_{j-2}^7 \qquad (19)$$

When considering $0 < Pe_{j-2} < 1$, we can write $\dfrac{Pe_j}{Pe_{j-2}^2} < \dfrac{Pe_{j-2}}{Pe_{j-2}^2}$. Consequently, $Pe_j < Pe_{j-1}$ can be obtained.

*…to memory of my brother Serdar Onur Alagöz.*